# Multi-Criteria Routing for Cryptographic Key Exchange in QKD Networks: Survey, Taxonomy, and a Conceptual Optimization Framework

Ivan Cvitić[1*] · Dragan Peraković[1] · Armando Nolasco Pinto[2,3]

[1] University of Zagreb, Faculty of Transport and Traffic Sciences, Vukelićeva 4, 10000 Zagreb, Croatia

[2] University of Aveiro, Department of Electronics, Telecommunications and Informatics (DETI), Campus Universitário de Santiago, P-3810-193 Aveiro, Portugal

[3] Instituto de Telecomunicações, Campus Universitário de Santiago, P-3810-193 Aveiro, Portugal

* Corresponding author: ivan.cvitic@fpz.unizg.hr

ORCID: I. Cvitić 0000-0003-3728-6711 · D. Peraković 0000-0002-0476-9373 · A. N. Pinto 0000-0003-2101-5896

Abstract Quantum Key Distribution (QKD) networks enable unconditionally secure key exchange using quantum mechanical principles; however, routing cryptographic keys across multi-hop quantum networks introduces challenges unique to quantum communication. This paper analyzes and classifies 26 routing strategies proposed between 2013 and 2024 for terrestrial, satellite, and hybrid QKD infrastructures, organizing them within a three-dimensional taxonomy spanning network type, optimization strategy, and routing objective. Dynamic, key-aware routing algorithms reduce service rejection rates relative to static shortest-path methods by incorporating real-time key-pool availability and link error rates; in the illustrative simulation we introduce in Section 6.7, key-aware overflow routing lowers the rejection rate by 25–40% across the practical quality-of-service operating range, consistent with the SDN-based dynamic-routing results reported by Bi et al. [15]. Multi-path strategies improve resilience against trusted-node compromise by distributing keys across disjoint routes, at the cost of increased key consumption; our simulation shows this overhead reaching 30–60% when roughly a quarter to a half of sessions adopt dual-path relay. SDN-based orchestration frameworks emerge as essential enablers of flexible, QoS-aware routing for hybrid networks integrating terrestrial fibers and intermittent satellite links. Building on the gaps identified in the surveyed literature, we propose a conceptual multi-criteria routing optimization framework that jointly models key delivery performance, trust exposure, resource cost, and resilience as competing objectives over a weighted graph representation of the QKD network, combining subjective and objective criteria weighting with adaptive link-state estimation suited for SDN-orchestrated hybrid infrastructures. Critical research directions include reducing dependence on trusted nodes through quantum repeaters or post-quantum cryptography, developing standardized key management interfaces for interoperability, and adopting AI-driven predictive routing for adapting to fluctuating network states. This synthesis of routing algorithms, performance trade-offs, and unresolved issues, together with the proposed framework, offers practical guidance for engineering secure, adaptable, and large-scale QKD networks capable of meeting real-world deployment demands.

## 1 Introduction

Quantum Key Distribution (QKD) provides a secure method for generating cryptographic keys, using quantum mechanical phenomena to ensure that any attempt at eavesdropping is detectable. Unlike classical key exchange protocols, QKD ensures information-theoretic security, making it a critical enabler for future-proof communications. With the development of terrestrial fiber-based and satellite-based QKD systems, networked QKD infrastructures are becoming a reality, requiring efficient and secure routing of quantum keys.

This paper is an extended version of our earlier survey presented at the 9th EAI International Conference on Future Access Enablers of Ubiquitous and Intelligent Infrastructures (FABULOUS 2025) [1]; the present work substantially deepens that survey and builds upon it to develop a formal multi-criteria routing optimization framework.

Classical public-key cryptosystems are under imminent threat from quantum computing, motivating the development of QKD as a means to distribute encryption keys with unconditional security. QKD leverages quantum mechanics (e.g., Fock states or coherent states) to generate shared random keys between two parties such that any eavesdropping on the quantum channel is detectable by increased error rates [2]. Over a point-to-point link, QKD can produce symmetric keys that are proven secure against even quantum adversaries. However, practical QKD links face severe distance limitations (on the order of tens to a few hundred kilometers in fiber) due to signal loss and detector noise. This necessitates QKD networks that connect multiple QKD links, enabling key exchange over long distances and among many users.

The most mature approach for extending QKD range is to use trusted intermediate nodes, where each fiber or free-space QKD link terminates at a node that performs key relay (by XOR-combining keys from adjacent links) to forward secret keys across the network. Trusted-node QKD networks have been demonstrated on city and national scales. For example, the Tokyo QKD network (2010) with 10 nodes [3] and the SECOQC Vienna network (2008) with 6 nodes [4] were early metropolitan QKD networks. More recently, a 46-node quantum metropolitan area network was implemented in China [5]. Going beyond terrestrial fiber, QKD-capable satellites in low Earth orbit (LEO) and geostationary orbit (GEO) have been used to establish intercontinental quantum links, culminating in an integrated space-to-ground QKD network spanning 4,600 km demonstrated in China [6]. These advances herald the dawn of global QKD networks that seamlessly blend terrestrial and satellite channels.

In parallel with this experimental progress, significant research attention has turned to the network-layer problem of routing keys in QKD networks, determining how keys should be relayed or shared across multiple hops to efficiently and securely reach their destinations. Unlike classical data routing, QKD key routing has unique challenges. First, QKD links continuously generate secret bits at a finite key generation rate that depends on channel loss and protocol efficiency. These keys are a consumable resource: if a path's intermediate node runs out of fresh key material, further secret communication on that path is blocked until more keys are generated. This makes key availability a critical factor in routing decisions.

Second, every additional hop (trusted relay) in a route introduces a potential security vulnerability, if an adversary compromises a trusted node, any keys passing through that node could be exposed. Routing strategies thus directly affect the threat model: a shorter route with fewer hops is generally more secure (fewer points of attack) but might suffer higher loss on a single long link, whereas multi-hop routes can achieve longer reach at the cost of trusting more nodes. Third, QKD networks often operate alongside classical networks, requiring hybrid routing approaches where a classical control plane (for signaling and key management) coordinates the usage of quantum channels. Issues such as network topology changes, node failures, and varying channel conditions (especially in free-space/satellite links subject to weather and orbital dynamics) demand dynamic routing algorithms beyond static paths.

The first objective of this work is to survey how optimal routes for key exchange in QKD networks can be determined, accounting for the above challenges. We cover the full spectrum of QKD network scenarios: (i) terrestrial QKD networks using optical fiber or line-of-sight free-space links with trusted repeater nodes; (ii) satellite QKD networks where keys are delivered via satellites acting as transmitters or relays; and (iii) hybrid networks that integrate terrestrial and satellite QKD links into a unified architecture. We review both theoretical models that formulate routing as an optimization problem (often with multi-objective criteria like maximizing key rate while minimizing hops or cost), as well as experimental and simulation-based studies that demonstrate routing and key relay in real or emulated QKD network testbeds. Particular emphasis is placed on classical-quantum hybrid routing approaches, for instance, using software-defined networking (SDN) controllers to manage QKD resources, or combining QKD with post-quantum cryptography for resilience. We also examine the security implications of different routing strategies, evaluating how well they uphold end-to-end key secrecy under various threat models. The survey compares routing algorithms on multiple performance axes, including achievable key throughput (keys/sec delivered between end-users), latency (time to establish keys), network resource utilization, and deployment cost.

The contribution of this paper is twofold. First, it provides a structured survey of routing approaches for cryptographic material exchange in QKD networks (covering 26 strategies published between 2013 and 2024), organized within an explicit taxonomy spanning network type, optimization strategy, and routing objective. Second, building on the gaps identified through this survey, it outlines a conceptual multi-criteria routing optimization framework, developed within the QRO-Route research project, that formulates key exchange routing as a constrained multi-objective optimization problem over a weighted graph representation of the network. Within this framework, routing is cast over four competing objectives, key-delivery performance, trust exposure, resource cost, and resilience, three of which (maximum security, minimum key-exchange time, and minimum cost) are formalized as single-objective programs that reduce to standard shortest-path computations, and are then unified through a hybrid weighting scheme that combines subjective (AHP) and objective (entropy) criteria weights together with adaptive link-state estimation, so that the framework integrates directly into the SDN orchestration loop. The work is guided by three research questions. RQ1: Which routing objectives, metrics, and optimization strategies have been proposed for QKD networks, and how can they be systematically classified? RQ2: How do existing approaches address the specific constraints of terrestrial, satellite, and hybrid QKD infrastructures, and which performance trade-offs do they entail? RQ3: Which requirements should a unified, multi-criteria routing framework satisfy in order to close the gaps identified in the literature, and how can such a framework be structured to remain compatible with SDN-orchestrated network control? The remainder of the paper is organized as follows: Section 2 positions this work with respect to previous research; Section 3 summarizes the characteristics of QKD network scenarios; Section 4 describes the review methodology; Section 5 analyzes the surveyed routing approaches; Section 6 introduces the proposed framework; Section 7 discusses trends and open challenges; and Section 8 concludes the paper.

# 2 Previous Research

Numerous routing algorithms for QKD networks have been proposed, each addressing the inherent limitations of quantum communication links. Early approaches aimed to abstract the quantum layer and utilize classical routing strategies over a virtualized infrastructure, for example via SDN-enabled overlay techniques [8]. Other works investigated adaptive and dynamic routing schemes that account for key buffer status and real-time link availability. Kutzkov and Kiktenko introduced a multiple non-overlapping paths routing algorithm to improve resilience in trusted-node networks by maintaining alternative disjoint routes [9]. This approach considers both key availability and link reliability (distributing secret keys across parallel paths to prevent single-point failure). Similarly, Lin et al. proposed reinforcement learning-based methods to dynamically select routes based on observed network behavior and key utilization patterns [10].

Routing in satellite-based QKD networks has gained traction with projects such as the European Space Agency's INT-UQKD initiative [7], which explores integrating LEO satellites into a global QKD framework. These efforts highlight the need for path planning that accommodates orbital dynamics, variable weather conditions, and the limited time windows for satellite visibility. Hybrid QKD networks, combining terrestrial fiber links with satellite relays, require sophisticated algorithms to balance key distribution across very different media. For example, heuristic approaches have been introduced to optimize hybrid routing for minimal latency and sufficient key throughput (e.g., by prioritizing either fiber or satellite paths based on current network conditions) [11]. The inclusion of quantum repeaters and memory-enabled nodes has also been considered in recent studies, which emphasize the impact of quantum memory decoherence on route selection and network performance (e.g., limiting the time keys can be stored before use).

More recent contributions have highlighted the need to integrate SDN controllers to manage hybrid quantum-classical infrastructures. Such architectures support modular network configuration and monitoring, which is essential for dynamic environments like satellite networks with time-varying connectivity. However, most existing proposals remain limited to conceptual frameworks or small-scale simulations, lacking practical validation in real-world quantum networks. Furthermore, few approaches systematically incorporate both dynamic key availability and security constraints (e.g., the trust level of intermediate nodes) into unified routing decisions. This gap underscores the need for integrated routing solutions that combine real-time control, quantum-aware metrics, and scalable orchestration across diverse QKD technologies, a challenge we address by synthesizing and evaluating strategies from the literature under a unified framework.

In relation to prior review and overview efforts, the present work differs in scope and intent. Existing syntheses of QKD networking predominantly concentrate on a single deployment scenario, most often terrestrial trusted-relay networks, or on a single architectural aspect, such as quality-of-service provisioning [17], standardization frameworks [26], or simulation tooling [28], [29], [30]. Comparative treatments that jointly consider terrestrial, satellite, and hybrid infrastructures under a common set of evaluation dimensions remain rare, and, to the best of our knowledge, none of the available reviews translates the identified shortcomings into a concrete set of design requirements for a next-generation routing mechanism. This paper addresses both points: it applies a uniform three-dimensional classification across all 26 surveyed strategies, and it operationalizes the resulting gap analysis in the form of a conceptual multi-criteria optimization framework (Section 6) intended to serve as a blueprint for subsequent formalization, implementation, and experimental validation.

# 3 Overview of QKD Network Characteristics

QKD networks comprise nodes capable of generating, storing, and forwarding quantum keys. The physical layer may include optical fibers, free-space terrestrial links, or satellite-to-ground channels. Unlike classical data, the secret key produced by a QKD link accumulates in a finite key pool at a time-varying rate set by the physical channel, and each secured session consumes key material from that pool in proportion to the amount of data protected — one key bit per message bit under one-time-pad encryption, or a smaller but non-zero amount under symmetric key-management schemes such as AES rekeying. A link whose pool is drained faster than it is replenished becomes temporarily unusable until fresh key is generated. This consumption-driven depletion, rather than any reuse limit on individual keys, is what necessitates continuous key renewal and resource-aware routing. Table 1 compares the typical parameters and constraints of three major QKD network scenarios: terrestrial, satellite, and hybrid.

**Table 1** Overview of terrestrial, satellite, and hybrid QKD networks [3], [4], [5], [6], [12]

| **Aspect** | **Terrestrial QKD Networks** | **Satellite QKD Networks** | **Hybrid QKD Networks** | **References** |
|---|---|---|---|---|
| Typical link distance | Up to ~100–200 km per fiber link (with trusted repeaters for longer distances). | Up to ~1200 km per satellite downlink (LEO to ground); GEO satellite links up to ~36,000 km. | Combines fiber segments (tens to hundreds of km) with satellite hops (hundreds to thousands of km). | [3], [5], [6], [12] |
| Key generation rate | Moderate to high on short fiber links (kilo- to megabits per second); decays exponentially with distance due to fiber loss (0.2 dB/km). | Lower (typically a few kilobits per second) due to channel loss and background noise; varies with elevation angle and weather conditions. | Highly variable; satellite links can overcome fiber loss over long distances (demonstrated with 1550 nm decoy-state BB84 QKD), but end-to-end key rate is limited by the weakest link in the path. | [5], [11] |
| Trust assumptions | Requires trusted intermediate nodes for multi-hop transmissions beyond the loss limit. | A satellite can be treated as a trusted node or as an untrusted relay depending on the protocol (e.g., entanglement-based schemes). | Combination of ground trusted nodes plus trust in the satellite (if used as a relay). This can reduce the number of intermediate trust points on the ground. | [9], [18] |
| Routing dynamics | Mostly static topology – links are stable, and routing can be optimized for a fixed infrastructure. | Dynamic topology (moving LEO satellites); requires time-dependent routing and scheduling of intermittent links. | Semi-dynamic: ground subnets are static, while satellite links are intermittent – adaptive routing is needed to choose between terrestrial and satellite paths. | [14], [20] |
| Key storage (buffering) | Key pools at nodes can buffer surplus keys to decouple generation and usage, smoothing out bursts. | Key storage on satellites and ground stations is critical due to intermittent connectivity (store-and-forward of keys between contacts). | Key pools at both ground nodes and satellites facilitate hybrid key relay between network segments. | [15], [20] |
| Main optimization goals | Maximize network key throughput; minimize number of hops (to limit exposure to potentially compromised nodes); balance load across links to avoid key depletion. | Schedule satellite passes to serve as many nodes as possible; optimize key usage before a link's outage; minimize wait times for key delivery. | Select optimal mix of terrestrial and satellite segments for each session – trading off latency, security, and cost (e.g., expensive satellite usage). | [15], [17], [21] |
| Notable demonstrations | SECOQC (Vienna) 6-node QKD mesh network [3]; Tokyo QKD 10-node network (ring topology) [2]; 46-node metropolitan QKD network in China [4]. | Micius satellite QKD downlink (China, ~1200 km); satellite-to-multiple-ground entanglement-based key distribution (Micius to two ground stations) [11]; Los Alamos/Oak Ridge (USA) utility QKD demo network (2019). | China integrated space–fiber QKD network (~4,600 km) [5]; EU EuroQCI quantum communication initiative (ongoing). | [3], [4], [5], [6] |

As shown in Table 1, QKD network scenarios range from relatively static fiber-based infrastructures to highly dynamic satellite constellations. Correspondingly, a variety of routing and key management strategies have been proposed to optimize key exchange under these differing conditions. This survey synthesizes these diverse approaches into a unified overview. To the best of our knowledge, it is among the first to jointly address terrestrial, satellite, and hybrid QKD network routing, supported by an extensive literature review spanning 2013–2024. In the following sections, we outline our methodology and then discuss routing strategies for each network type in detail.

# 4 Research Methodology

To classify and compare existing routing strategies, we conducted a structured literature review of peer-reviewed publications, technical reports, and major QKD deployment initiatives. The selected literature spans the period from foundational QKD networking experiments (e.g., the DARPA Quantum Network in the mid-2000s and SECOQC in 2008) to contemporary hybrid network designs employing SDN-based control and satellite integration. The research selection criteria included:

- Relevance to QKD-specific constraints: e.g., key exhaustion, probabilistic key generation, trust management.
- Applicability to trusted-node architectures: multi-hop key relay.
- Consideration of physical-layer limitations: loss, noise, and topology dynamics.
- Support for dynamic or adaptive routing mechanisms.
- Experimental validation or realistic simulation.

The review followed the PRISMA 2020 guidelines for identification, screening, and inclusion. Candidate records were retrieved from five bibliographic databases — Web of Science, Scopus, IEEE Xplore, the ACM Digital Library, and arXiv — using the Boolean query "quantum key distribution" AND ("routing" OR "path selection" OR "key relay" OR "scheduling") applied to titles, abstracts, and keywords, restricted to publications from January 2013 to December 2024. The database search was complemented by backward and forward snowballing from the reference lists of retrieved publications and by gray literature describing major deployment initiatives (SECOQC, the Tokyo QKD Network, the Chinese integrated space-to-ground network, and EuroQCI).

The search returned 312 records. After removal of 98 duplicates, 214 records were screened by title and abstract against the criteria above; 148 were excluded as out of scope, leaving 66 reports assessed in full text. Of these, 35 were retained as primary and supporting sources, which collectively describe the 26 distinct routing strategies analyzed in detail, together with the foundational experimental deployments, standardization documents, and simulation-platform studies used as supporting context. Records were excluded when they (i) did not address routing, path selection, key relay, or scheduling in a QKD context; (ii) considered only point-to-point QKD links with no multi-hop or network dimension; or (iii) were superseded by a later, more complete version by the same authors. These counts reflect the search as executed in December 2024 and may vary with search date and database access.

Each approach was analyzed along three dimensions: (1) Routing objective (throughput maximization, latency minimization, reliability enhancement, etc.), (2) Network type considered (terrestrial, free-space, satellite, or hybrid), and (3) Optimization strategy (static vs. dynamic, deterministic vs. probabilistic, heuristic vs. machine learning). We further assessed whether the solutions support integration with classical control protocols, SDN controllers, or centralized key management systems. This methodology facilitated the construction of a comparative framework summarizing the strengths, limitations, and gaps in current QKD routing proposals.

# 5 Analysis of QKD Routing Approaches

## 5.1 Taxonomy of Surveyed Approaches

The 26 surveyed routing strategies can be positioned within a three-dimensional classification space derived from the analysis dimensions introduced in Section 4. The first axis, network type, distinguishes terrestrial fiber-based networks, satellite (free-space) networks, and hybrid infrastructures combining both. The second axis, optimization strategy, separates static deterministic schemes (classical shortest-path variants with QKD-aware edge weights), dynamic heuristic schemes (greedy, load-balancing, and residual-key-aware approaches with periodic recomputation), stochastic and probabilistic models (e.g., Poisson-based blocking probability estimation), mathematical programming formulations (integer linear programming for joint routing, scheduling, and resource assignment), and learning-based methods (reinforcement learning and emerging AI-driven prediction). The third axis, routing objective, captures the primary optimization goal: key throughput maximization, latency minimization, trust exposure minimization, resilience and availability maximization, quality-of-service differentiation, or deployment and operational cost reduction.

Projecting the surveyed literature onto this space reveals a markedly uneven distribution. The majority of contributions target terrestrial trusted-relay networks with dynamic heuristic strategies optimizing a single primary objective, most often key throughput or service blocking probability, while security is typically treated only implicitly, through hop-count minimization rather than as an explicit, quantified objective. Mathematical programming formulations are concentrated in the satellite and optical-backbone domains, where contact scheduling and wavelength assignment naturally lend themselves to ILP modeling, but their computational cost limits applicability to online routing decisions. Learning-based approaches remain exploratory, with validation restricted to small simulated topologies. Most notably, only a small minority of the surveyed strategies jointly consider three or more objectives within a single decision model, and none combines explicit trust quantification, key availability dynamics, resource cost, and resilience under a common weighting mechanism. This observation directly motivates the framework proposed in Section 6.

## 5.2 Terrestrial QKD Networks

In fiber-based QKD networks, routing must contend with high attenuation over long distances. Trusted nodes extend the reach by storing and relaying keys between shorter fiber segments. Algorithms such as those proposed in research [21], use dynamic path selection based on key buffer availability. Multi-path strategies, as in [9], enhance robustness by maintaining alternate disjoint routes to avoid single points of failure.

**Table 2** Routing algorithms in terrestrial QKD networks

| **Approach** | **Key Metric** | **Reference** |
|---|---|---|
| Dijkstra-based routing | Key cost per hop (link weight) | [13] |
| Residual key-aware shortest path | Available key material on links | [14] |
| Blocking probability estimation | Poisson-based link stability | [15] |
| Multi-path XOR-based routing | Disjoint-path security; throughput | [9] |
| Reinforcement learning (Q-learning) | Adaptive reward for successful key delivery | [10] |

Terrestrial QKD networks, primarily relying on optical fiber infrastructure, are currently the most mature and widely deployed form of quantum key distribution. Their routing strategies, however, must address unique constraints arising from limited key generation rates, sensitivity to quantum bit error rate (QBER), and the necessity of trusting intermediate nodes as seen in Table 2. A critical limitation of terrestrial QKD links lies in their attenuation and noise characteristics, with secret key rates decreasing exponentially over distance due to fiber loss (~0.2 dB/km) and dark counts at detectors. As a result, trusted repeater nodes are introduced to extend reach, but every additional node introduces a potential compromise point, shifting routing optimization to a multi-objective problem: minimize hops for trust, maximize key throughput, and avoid overloading individual links.

Another unique aspect is the key resource constraint: QKD links produce a consumable key material. This resource constraint makes it essential for routing protocols to monitor the current state of key buffers and adapt accordingly to avoid service disruptions due to key depletion.

Recent approaches extend classical shortest-path algorithms to account for quantum-specific metrics. For example, in residual-key-aware routing, each edge weight is dynamically computed based on the availability of key material and the link's expected QBER.

In [15], authors model key generation and consumption using a Poisson process, calculating the blocking probability of each link and adjusting weights accordingly. This SDN-based dynamic routing strategy has been shown to significantly reduce service rejection and latency compared to static OSPF-based methods.

To ensure route feasibility under limited key availability, links falling below a predefined key threshold can be excluded from the routing graph entirely, necessitating frequent state synchronization between nodes and SDN controllers or distributed agents [13]. This approach guarantees that selected paths meet minimum operational criteria for secure communication. The methodology has been further enhanced by integrating QBER thresholds directly into edge weights, thus preemptively avoiding high-error links that are likely to fail during key reconciliation [13].

As QKD networks are increasingly deployed in shared metro-area environments, support for multiple tenants and differentiated QoS requirements has become essential. For instance, real-time applications such as encrypted voice calls demand consistent key delivery, whereas bulk encrypted transfers can tolerate temporal fluctuations. To accommodate these needs, utility-based routing strategies have been proposed, wherein key bandwidth is allocated according to service-level agreements (SLAs) [17], [18]. These methods utilize SDN controllers to assign and enforce service priorities through routing and scheduling mechanisms. Consequently, high-priority sessions maintain uninterrupted access to key material, while lower-priority sessions may be deferred or rerouted during key scarcity.

Such QoS-aware routing is facilitated by SDN-based key orchestration frameworks that virtually slice key resources among tenants. Policy-driven mechanisms define how and when each tenant can access the underlying quantum infrastructure, ensuring both fairness and efficiency in resource utilization.

Load-balancing mechanisms can mitigate key exhaustion by distributing traffic across multiple paths. However, this increases coordination complexity. For example, Wang et al. suggested periodically recalculating optimal paths based on real-time key usage metrics. Routing decisions in terrestrial QKD networks must also consider fiber quality (loss), the availability of dark fibers, and the geographic distribution of trusted relay stations. Notably, the SECOQC Vienna QKD network implemented a dynamic routing protocol to handle variations in link conditions and key supply.

### 5.3 Satellite QKD Networks

Satellite QKD introduces a unique set of constraints compared to terrestrial fiber-based networks. Unlike stable terrestrial links, satellite links are highly time-dependent, intermittent, and asymmetric. This results in a routing and scheduling challenge rather than a straightforward pathfinding problem.

Low Earth Orbit (LEO) satellites exhibit periodic visibility windows with ground stations, producing a dynamic network topology best represented by time-expanded graphs, where each edge corresponds to a satellite-ground contact opportunity during a specific time slot. A topology abstraction method was introduced to construct a static logical layer that aggregates all feasible satellite-ground pairings over a given pass interval, allowing the application of classical routing algorithms such as Dijkstra or ILP for determining optimal contact sequences for key relaying [20].

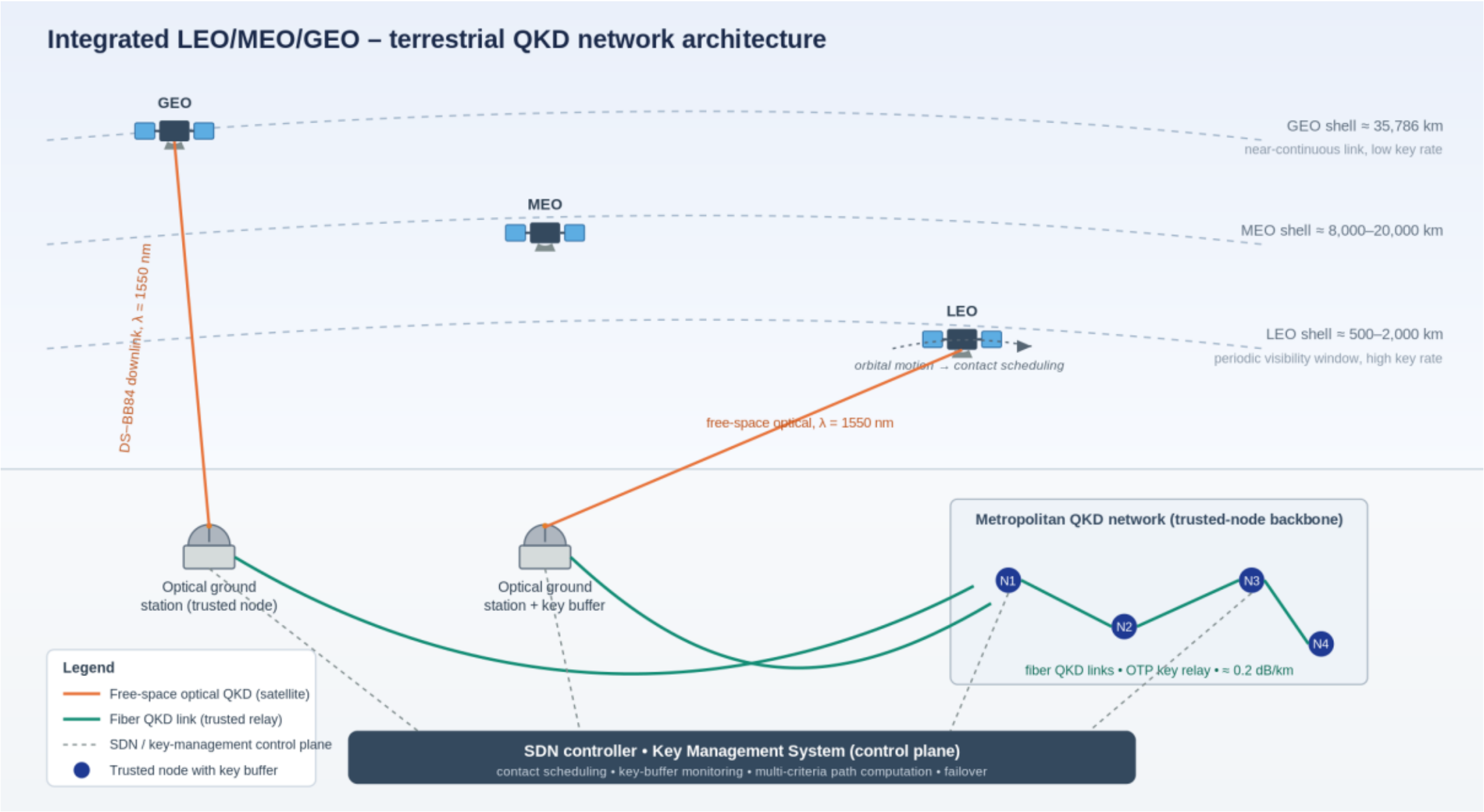

**Fig. 1** Conceptual architecture of an integrated LEO/MEO/GEO–terrestrial QKD network, showing free-space optical downlinks (DS–BB84, λ = 1550 nm), a trusted-node fiber backbone with OTP key relay, and an SDN/key-management control plane

In contrast, Geostationary Earth Orbit (GEO) satellites provide near-continuous connectivity but suffer from significantly lower key rates due to increased optical losses. Thus, routing strategies in hybrid LEO-GEO networks must balance link availability, throughput, and latency. An example of the envisioned architecture integrating urban and terrestrial grid networks across LEO, MEO, and GEO orbits for DS-BB84 QKD downlinks under night-time conditions is shown in the schematics in Figure 1.

Satellite QKD requires solving the contact scheduling problem, determining which users or links are to be served during each satellite pass. An ILP framework has been proposed to maximize key delivery while adhering to constraints imposed by orbital dynamics and visibility windows [24]. To improve scalability, heuristic methods, including greedy and hybrid metaheuristic approaches like genetic algorithms, have been developed for resource allocation under service-level guarantees.

Another critical design consideration is key storage. Due to the intermittent nature of satellite links, generated keys must be temporarily stored onboard and forwarded during subsequent contact opportunities. Therefore, buffer management strategies and time-slot reservation mechanisms are essential components of satellite QKD system design. Furthermore, environmental factors such as cloud cover, beam divergence, and atmospheric turbulence influence link availability and quality. Real-time environmental data integration and predictive weather modeling are increasingly being explored for adaptive satellite QKD routing.

## 5.4 Hybrid QKD Networks

Hybrid QKD networks combine terrestrial and satellite components. Routing in such systems involves selecting among heterogeneous paths with different trust levels and performance profiles. For example, researchers showed a hybrid implementation that dynamically switches between satellite and fiber links in a large-scale network, shown in Figure 2 [6].

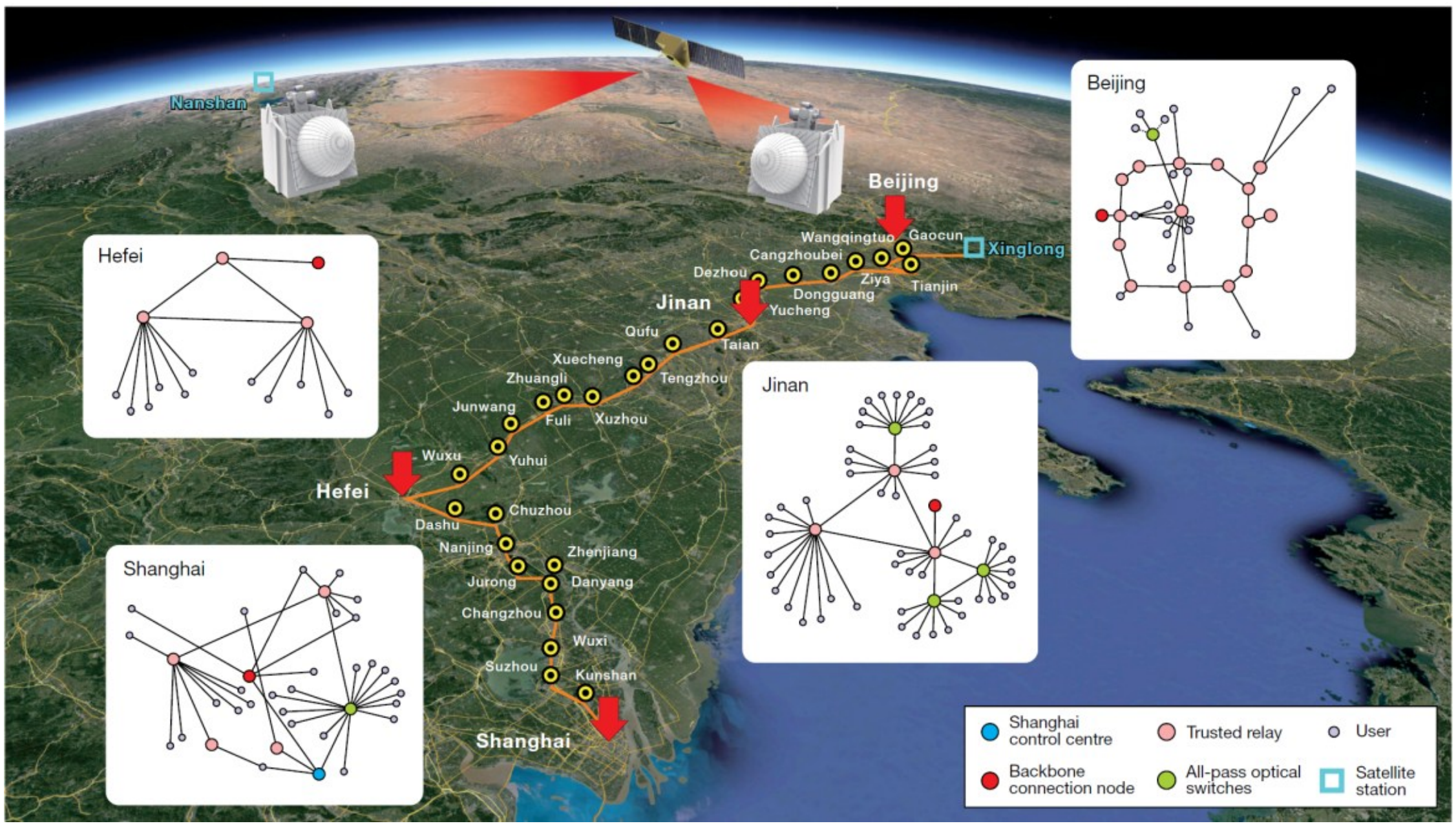

**Fig. 2** Visualization of integrated space-to-ground quantum network [6]

Heuristic approaches have been proposed to minimize total latency while ensuring sufficient key material is available for end-users. Other studies consider the use of quantum repeaters and quantum memory, which store quantum states (entangled photons and their quantum information) rather than the sifted classical key bits; this is distinct from the classical buffering of already-measured key material in conventional memory at a node. Quantum memory can enable delayed entanglement swapping to match link availability and, in principle, removes the need to trust intermediate nodes, but practical quantum memory with sufficient storage time and fidelity remains an open experimental challenge. In general, hybrid routing algorithms must balance trade-offs between performance (throughput, delay), trust (number of intermediaries), and cost. For instance, a routing decision might favor a longer terrestrial path over a satellite hop at certain times to avoid using scarce satellite resources or to maintain a higher security level, and vice versa during off-peak hours.

## 5.5 Routing Algorithms and Protocols

Early routing strategies in QKD networks adapted classical algorithms like Dijkstra's shortest path, with edge weights reflecting QKD-specific metrics such as key rate or trust level. A dynamic graph approach was proposed in which only links with available keys are considered during route computation [13]. It was also shown that standard protocols like OSPF, when applied without modification, can lead to the overuse and depletion of certain QKD links; to address this, probabilistic link-cost metrics based on key consumption rates, such as Poisson-based models for key exhaustion, were introduced to improve routing decisions [15]. Recent developments have focused on SDN-controlled dynamic routing approaches, wherein link weights are updated in real time based on current key buffer levels and predicted key lifetimes. For instance, adaptive link weighting and time-sensitive key stability models have been proposed to optimize path selection [14], [15], where the process of a key pool providing key encryption for data communication between two optical nodes can be seen in Figure 3.

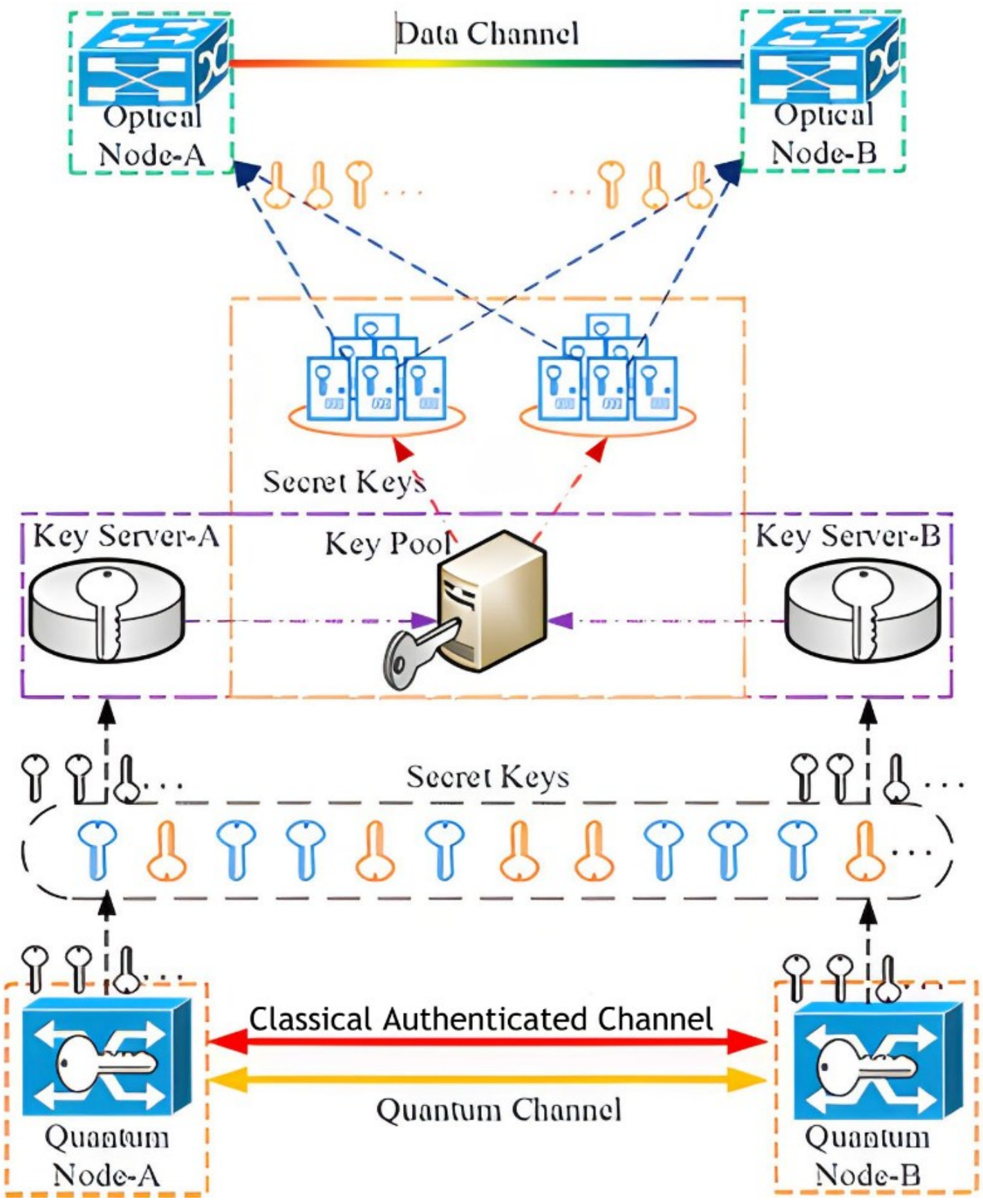

**Fig. 3** Quantum key pool [14]

Multi-path routing has been explored as a strategy to enhance robustness and confidentiality: one approach employed XOR-based schemes to distribute secret keys across disjoint paths, which reduces the risk from compromised trusted nodes at the cost of requiring multiple simultaneous key-distribution routes [9]. This concept was further developed for increased throughput using parallel fiber links in a metropolitan QKD deployment [16].

**Table 3** Comparison of QKD routing strategies

| Routing Strategy | Key Features | Advantages | Limitations | Reference |
|---|---|---|---|---|
| Shortest-Path (Key-Aware) | Minimizes hop count or maximizes key rate | Simple, widely supported | Ignores key depletion risk | [13] |
| Greedy Routing | Quick selection with limited information | Fast, scalable | May cause link overloading | [27] |
| Residual Key-Based | Considers remaining key pools | Improved reliability under key constraints | Needs frequent updates | [14] |
| Dynamic (SDN-enabled) | Monitors key state and adapts in real-time | Responsive to network dynamics | Requires centralized intelligence | [15] |
| Reinforcement Learning | Learns optimal paths via feedback/reward | Handles complex environments | High computational overhead | [10] |
| Multi-Path (XOR/Secret Sharing) | Splits keys across disjoint routes for security | Improves robustness against compromise | Consumes more keys | [9] |
| QoS-Aware Routing | Prioritizes sessions with guaranteed key rates | Enables service-level differentiation | Needs traffic classification | [17] |
| Multi-Tenant Partitioning | Sliced QKD resources among isolated tenants | Enables resource sharing and isolation | Complex policy management | [18] |

Quality-of-service (QoS)-aware routing is becoming increasingly relevant for multi-tenant QKD infrastructures. One solution employed a utility-based optimization model that prioritizes key delivery for high-importance users in the context of limited key availability [17]. Another contribution implemented SDN slicing to support tenant-isolated key provisioning in hybrid QKD networks, ensuring minimal interference across different user domains [18]. On the protocol development side, a dedicated link-state routing protocol was created specifically for QKD environments under the SECOQC initiative [4], while integration of QKD metrics into legacy routing protocols like OSPF has been experimentally validated in real-world testbeds [19]. Centralized SDN-based controllers continue to be favored in experimental setups due to their capacity for global state awareness and advanced optimization capabilities. Details can be seen in Table 3 and Table 4.

**Table 4** Study-by-study comparison of surveyed QKD routing approaches.

| Study | Network type | Routing metric | Optimization method | Trust model | Validation method | Performance metrics | Principal limitations |
|---|---|---|---|---|---|---|---|
| Han et al. [13] | Terrestrial (fiber) | Key cost per hop / optical-path switching | Dijkstra shortest-path on key-weighted graph | Fully trusted relays | Numerical analysis | Path key cost, connection success | Static weights ignore real-time key depletion |
| Lin et al. [14] | Terrestrial (ring) | Residual key material on links | Residual-key-aware shortest path; key-pool model | Partially-trusted ring nodes | Simulation | Key utilization, blocking | Ring-specific; limited topology generality |
| Bi et al. [15] | Terrestrial (virtual QKD net) | Link blocking probability (Poisson key model) | Dynamic SDN routing with adjustable key pools | Trusted relays under SDN control | Simulation | Service-rejection rate, latency, key-usage efficiency | Centralized controller; reactive to depletion |
| Kutzkov & Kiktenko [9] | Terrestrial (mesh) | Disjoint-path count / security | Multiple non-overlapping (XOR) path routing | Reduced trust via key splitting | Numerical analysis | Compromise resilience, throughput | Higher key consumption; needs disjoint paths |
| Yan et al. [10] | Terrestrial | Learned reward for successful key delivery | Reinforcement learning (Q-learning) | Trusted relays | Simulation | Key-delivery success rate | High training/computation overhead |
| Yu et al. [16] | Terrestrial (metro, parallel fiber) | Aggregate secret-key throughput | Collaborative multi-path secret-key provisioning | Partially-trusted relays | Simulation on optical network | Secret-key provisioning rate | Requires parallel disjoint fiber capacity |
| Mehic et al. [17] | Terrestrial (trusted-relay) | QoS / user-priority utility | Utility-based QoS-aware optimization | Trusted relays | Simulation | QoS satisfaction, prioritized key delivery | Needs traffic classification/policy |
| Cao et al. [18] | Terrestrial (backbone) | Trusted/untrusted relay mix; slicing | Hybrid trusted/untrusted relay optimization | Mixed trusted/untrusted relays | Simulation on backbone topology | Cost, security level | Complex policy; untrusted-relay hardware assumed |
| Guo et al. [20] | Hybrid GEO/LEO satellite | Time-extended link abstraction | Topology-abstraction routing over pass windows | Trusted ground + satellite nodes | Simulation | Secret-key provisioning, wait time | Depends on accurate pass scheduling |
| Zhang et al. [21] | Terrestrial (optical) | Routing, channel, key-rate, time-slot | Joint RCKTA resource assignment | Trusted relays | Optimization / simulation | Resource utilization, blocking | Combinatorial complexity at scale |
| Aguado et al. [22] | Hybrid (SDN/NFV) | Key-buffer level; failover | SDN/NFV service automation, adjustable key pool | Trusted relays under SDN | Field demonstration | End-to-end encryption continuity | Demonstrator scale; vendor-specific |
| van Duijn et al. [27] | Terrestrial (trusted relay) | Resource-aware path cost | Comparison of trusted-relay routing strategies | Trusted relays | Simulation | Key-distribution success vs. overhead | Higher success trades off compute overhead |
| Jia et al. [33] | Terrestrial (key-pool overlay) | Key-pool occupancy / cost | Cost-optimization over quantum key pool | Trusted relays | Optimization / simulation | Key-pool cost, blocking | Overlay assumptions; parameter sensitivity |
| Luo & Li [34] | Terrestrial (trusted relay) | Trust / causal consistency | Trust-aware causal-consistency routing | Adversarial / malicious nodes | Simulation | Resilience to malicious nodes | Added protocol overhead; trust bootstrap |
| Martin et al. (MadQCI) [31] | Hybrid (SDN-QKD, production) | QKD+PQC key provisioning | Heterogeneous SDN-QKD orchestration | Trusted relays + PQC hybrid | Production-network deployment | Scalability, interoperability | Integration complexity; early-stage standards |

## 5.6 Satellite, Hybrid, and Classical-Quantum Integration

Satellite QKD routing requires temporal optimization due to the inherently time-varying topology of orbital networks. One approach to addressing this challenge involves topology abstraction, which converts dynamic link states into quasi-static graphs, thereby simplifying routing decisions over orbital cycles [20]. Optimization methodologies, particularly those based on integer linear programming (ILP), have been employed to schedule key transmissions and allocate network resources efficiently. ILP-based models have been proposed for joint optimization of routing, channel assignment, and key rate allocation, demonstrating tangible benefits of global optimization when physical constraints are considered [21]. In parallel, the concept of hybrid QKD routing, which combines both satellite and terrestrial fiber segments, has gained traction. Software-defined networking (SDN) orchestration has been used to manage such hybrid topologies, enabling centralized control over mixed infrastructures [18], [22]. Adaptive routing strategies have been proposed for global-scale QKD deployments, integrating considerations such as channel loss and trust levels across fiber and satellite segments [23]. Importantly, the optimal routing strategy is not static but varies with environmental and operational conditions, for instance, satellites may be more effective during nighttime or clear weather, while fiber may be preferable in adverse conditions or when satellites are unavailable. SDN-enabled control planes support real-time adaptability, making latency-aware QKD routing with dynamic key refresh mechanisms feasible in practice [19], [24]. Moreover, hybrid encryption schemes that combine QKD with post-quantum cryptographic (PQC) techniques are being actively explored. Such schemes ensure continuity of secure communication by using QKD-generated keys when available, and falling back to PQC when quantum channels are degraded [7], [25]. Recent trials and projects have started validating these architectures, which represent a promising direction for enhancing the robustness of secure communication infrastructures.

In modern QKD networks, a centralized SDN controller (control layer) orchestrates both the quantum and classical resources. It maintains a global view of the network: querying each QKD link's status (available key buffer, error rates) and each node's classical switches or encryptors. The controller uses this information to schedule key generation and path setup for user requests. For example, it can reserve an optical circuit for data while ensuring sufficient QKD-generated keys are routed along that path. The SDN controller "sees" QKD devices as network elements that export their capabilities (without revealing secret keys), and can command them to start/stop key sessions or adjust rates. In effect, the SDN control plane unifies classical and quantum routing: it computes routes that optimize key throughput or latency based on real-time key-buffer metrics, then programs both the classical switches and the QKD hardware accordingly. It can also enact higher-level functions like traffic engineering or multi-tenancy: for instance, dynamically slicing the network so that each user's key demands are isolated [31], [15].

Routing decision flow (example): When an application requests a key between nodes, the SDN controller follows a loop of sensing, planning, and acting. A typical sequence can be seen below:

- Step 1: Key request & requirement. An application or user at node A requests a key of a certain length to node B. The requirement (e.g. length, deadline) is passed to the SDN controller.
- Step 2: Metrics collection. The controller polls the network elements: it reads each link's current key-buffer level and status (via a key-pool management interface), and checks classical link metrics (latency, availability).
- Step 3: Path computation. Using these metrics, the controller computes the best route(s). For instance, it may run a weighted shortest-path algorithm where each edge weight penalizes links with low key availability or high latency. Links with insufficient stored keys are either avoided or assigned very high cost. Multi-path or XOR-routing may be employed to split keys across diverse routes for robustness.

- Step 4: Route installation. The controller configures the network: it instructs the chosen QKD links to generate keys (perhaps boosting their rate) and programs classical switches to establish the data/management channels. In hybrid scenarios, it may allocate a satellite ground station or a fiber route as appropriate.
- Step 5: Monitoring & feedback. As keys are consumed, the controller continuously monitors key-buffer levels. If a buffer drops below a threshold, this triggers a feedback loop: the controller can recompute routes or request extra QKD generation on that link. For example, research [22] demonstrated an "adjustable key pool" mechanism where remaining key count is used to avoid routes that risk depletion.
- Step 6: Failover (reconfiguration). If a link fails or a buffer is critically low, the controller immediately activates a failover: it diverts the key flow to pre-computed backup paths (e.g. alternate fiber links or pending satellite passes). This might involve tearing down one path and instating another within milliseconds, or gradually shifting key traffic. The controller ensures that the end-to-end key service continues seamlessly despite link outages or depletions.

The above loop illustrates how SDN provides both agility and reliability: by tightly coupling control software with network measurements, routing decisions can adapt in real-time to key-generation dynamics and link conditions.

In hybrid fiber–satellite networks, the SDN controller handles even more complex scheduling. Satellite-based QKD links are intermittent and scheduled (known pass windows and weather constraints), whereas fiber links are relatively stable. The controller integrates these into its topology model. For example, it may reserve fiber key-buffer resources to "bridge the gap" between satellite passes, or it may schedule a ground station's QKD system to run during a predicted clear-sky window. Using abstractions the controller treats each satellite pass as a time-extended link in the graph [20]. It can then compute routes that combine multiple fiber segments and satellite hops, trading off key rates and latency. If a satellite link becomes unavailable (e.g. passing out of visibility or due to clouds), the SDN feedback loop immediately re-routes keys via terrestrial paths until the next satellite opportunity. Hybrid demonstrators have shown that a centralized SDN controller can effectively coordinate such mixed topologies pre-loading keys on ground links before a satellite relay is activated, and switching over as needed [22]. In this way, SDN orchestration ensures end-to-end key delivery across fiber and free-space segments, leveraging the best of each while responding automatically to the network's evolving state.

# 6 Towards a Multi-Criteria Routing Optimization Framework

The analysis in Section 5 shows that, while individual routing proposals successfully address isolated aspects of the QKD routing problem - key availability, trust exposure, contact scheduling, or service differentiation - no surveyed approach treats these aspects jointly within a single, technology-agnostic decision model. This section outlines the conceptual structure of a multi-criteria routing optimization framework for cryptographic material exchange, currently under development within the QRO-Route research project. The intent here is not to present a complete algorithm with experimental results, but to define the skeleton of such an algorithm: its network model, objective space, constraint set, weighting mechanism, and control-plane integration. In doing so, the framework operationalizes the gap analysis of this survey into a set of verifiable design decisions that can subsequently be formalized, implemented, and evaluated.

### 6.1 Motivation and Design Principles

Six design principles are derived directly from the shortcomings identified in the surveyed literature. P1: Joint treatment of key availability and security, key buffer dynamics and trust exposure must enter the same decision model rather than being optimized in isolation. P2: Technology-agnostic abstraction, the model must represent fiber, free-space, and satellite links uniformly, so that hybrid paths can be compared on equal terms. P3: Explicit multi-objective formulation - performance, security, cost, and resilience should be modeled as separate, individually measurable objective functions rather than collapsed prematurely into a single ad hoc composite metric. P4: Separation of policy and state, subjective operator priorities (e.g., the relative importance of security versus latency for a given service class) must be decoupled from objective, measured network state, with each contributing its own weighting component. P5: SDN compatibility, the framework must map naturally onto the sensing–planning–acting control loop described in Section 5.6, reusing standardized key management and control interfaces. P6: Computational tractability, the online routing decision must remain solvable at request arrival timescales, delegating heavyweight optimization to offline or periodic computation.

### 6.2 Network Model

The QKD network is modeled as a weighted graph G = (V, E, W). The node set V is partitioned into end nodes (key consumers), trusted relay nodes, and, where applicable, untrusted intermediary nodes (e.g., measurement-device-independent relays or satellites operated in untrusted mode). Each edge $e \in E$ represents a key-generating link of arbitrary underlying technology - optical fiber, terrestrial free-space, or satellite-to-ground - and is annotated with a weight vector $w(e) = (R_e, Q_e, B_e, T_e, C_e, A_e)$, whose components capture, respectively: the current secret key generation rate, the measured quantum bit error rate, the residual key buffer occupancy, a trust coefficient associated with the link's terminating relay, a normalized resource cost (reflecting, e.g., scarce satellite contact time or leased dark fiber), and link availability. For stable fiber links, availability is effectively binary; for satellite links, it is expressed through time windows, allowing the time-expanded graph technique surveyed in Section 5.3 [20] to be embedded directly into the same model. A routing request is defined by a source–destination pair, a required key volume, a deadline, and a service class. The routing decision then consists of selecting a path - or, in the multi-path variant, a set of node-disjoint paths - through G that optimizes the objective functions defined below, subject to the constraint set in Section 6.4.

### 6.3 Objective Functions

The framework defines four objective functions evaluated over a candidate path P. F1 (performance) maximizes the expected key delivery capability of the path, dominated by its bottleneck: the minimum, over all links of P, of the effective key supply (residual buffer plus expected generation within the request deadline). F2 (security) minimizes the aggregated trust exposure of the path, expressed as a function of the number of trusted relays traversed and their individual trust coefficients, so that a path through fewer or more strongly protected relays is explicitly preferred, generalizing the implicit hop-count minimization found in the surveyed literature into a quantified objective. F3 (cost) minimizes the cumulative normalized resource cost of the path, capturing operational asymmetries such as the premium attached to satellite contact time or to reserved high-priority key bandwidth. F4 (resilience) captures how well a routing decision withstands link failures and trusted-node compromise. It takes two forms, depending on the routing mode. In single-path routing it penalizes links whose recent behavior is unstable, those with a high variance of key rate or QBER, because such links are the most likely to drop below the usable key threshold and break the path. In multi-path routing it instead rewards spreading the key across several node-disjoint paths, so that the failure or compromise of any single trusted node cannot bring down the overall key exchange. These objectives are mutually conflicting by construction: the most secure path is rarely the cheapest, and the highest-throughput path may concentrate load and reduce resilience. Of these four objectives, three, security, key-exchange time (the operational form of the performance objective F1), and cost, admit exact single-objective formulations that reduce to shortest-path problems and are developed below as Scenarios S1–S3. The resilience objective F4, being defined over a set of paths rather than a single path, is not cast as a standalone scenario but is incorporated through the multi-path extension and the weighting of Section 6.5. The baseline formulation scalarizes the four objectives into a single path cost through the weighting mechanism of Section 6.5, enabling the use of efficient shortest-path computation online; a Pareto-front analysis of the unscalarized problem is foreseen as an offline tool for policy calibration.

To make these objectives precise we fix the following notation. A routing request is the tuple $\rho = (s, d, \kappa, \delta, c)$, with source $s \in V$, destination $d \in V$, required secret-key volume $\kappa$ (in bits), deadline $\delta$ (in seconds), and service class c. A candidate path P is an ordered sequence of links $e \in E$ connecting s to d, and $\mathscr{R}(P) \subseteq V$ denotes the set of intermediate trusted relays it traverses. The link weight vector $w(e) = (R_e, Q_e, B_e, T_e, C_e, A_e)$ of Section 6.2 supplies, per link, the secret-key generation rate $R_e$ (bit/s), the QBER $Q_e$, the residual buffer $B_e$ (bits), the trust coefficient $T_v \in (0, 1]$ of the relay v terminating the link (with $T_v \to 1$ denoting a fully hardened relay), the normalized resource cost $C_e$, and the availability $A_e$. The feasible set $\mathcal{P}(\rho)$ is the subset of all s–d paths that satisfy the hard constraints C1–C8 of Section 6.4; in particular, pre-filtering by C1 (residual key) and C2 (QBER limit) removes infeasible links and yields the pruned graph G′ on which all three scenarios below are solved. Each scenario is an instance of a constrained single-objective shortest-path program over G′; the scenarios differ only in the objective optimized, and therefore expose, at the extremes, the conflict that the weighting mechanism of Section 6.5 later mediates. The three canonical scenarios correspond directly to the three operator intents most frequently invoked in the surveyed literature: maximum security (S1), minimum key-exchange time / maximum key rate (S2), and minimum cost (S3).

**Scenario S1, maximum security (minimum trust exposure).** In a trusted-relay network the dominant residual security risk is the exposure of cleartext key at each intermediate relay; a path is only as secure as the weakest relay it must trust. Modeling relay-compromise events as independent, the probability that none of the traversed relays is compromised, the end-to-end trust level of the path, is $\mathrm{Sec}(P) = \prod$ over $v \in \mathscr{R}(P)$ of $T_v$. Maximizing this product is equivalent to minimizing its negative log-likelihood, so the security objective is the additive function

$$F_sec(P) = -\Sigma_\{v \in \mathscr{R}(P)\}\ \ln T_v\,, \qquad (1)$$

and Scenario S1 is the program

$$min_\{P \in \mathcal{P}(\rho)\}\ F_sec(P) = min_\{P \in \mathcal{P}(\rho)\}\ (-\Sigma_\{v \in \mathscr{R}(P)\}\ \ln T_v)\,. \qquad (2)$$

Because the per-relay term $w_v = -\ln T_v \geq 0$ is non-negative and additive, (2) is an ordinary shortest-path problem on G′ with relay weights $w_v$ and is solved exactly in $O(|E| + |V| \log|V|)$ by Dijkstra's algorithm. Two limiting cases are instructive. When all relays share a common trust level $T_v = t$, $F_sec(P) = |\mathscr{R}(P)|\cdot(-\ln t)$, so S1 reduces exactly to minimizing the number of trusted relays - the implicit hop-count heuristic of the surveyed literature [14], [17] - which is thereby recovered as a special case of (2). When relays are heterogeneously hardened, (2) departs from hop-count minimization and may prefer a longer path through strongly trusted nodes, consistent with the trust-aware path selection of [34] and the security-driven resource allocation of [37].

**Scenario S2, minimum key-exchange time (maximum key rate).** In hop-by-hop one-time-pad relaying the end-to-end secret-key throughput of a path is governed by its bottleneck link, since no relay can forward key faster than its slowest incoming link can supply it. A link e can immediately contribute its residual buffer $B_e$ and thereafter generate key at rate $R_e$, so the time for link e to furnish the requested volume $\kappa$ is $t_e(\kappa) = \max(0, (\kappa - B_e)/R_e)$. Adding the cumulative propagation and relay-processing latency $\ell_e$ of the traversed links, the end-to-end key-exchange time of a path is the min–max (bottleneck) quantity

$$F_time(P) = \max_{e \in P} \max(0, (\kappa - B_e)/R_e) + \Sigma_{e \in P}\, \ell_e\,, \qquad (3)$$

and Scenario S2 is

$$\min_{P \in \mathcal{P}(\rho)} F_time(P)\,. \qquad (4)$$

Equivalently, neglecting the typically small latency term, S2 maximizes the effective bottleneck key rate $R^{eff}(P) = \min_{e \in P}(B_e/\delta + R_e)$; it is therefore a widest-path (maximum-bottleneck-capacity) problem, solvable by a modified Dijkstra or Bellman–Ford in $O(|E| \log|V|)$. The deadline constraint C4 enters as the feasibility test $F_time(P) \le \delta$. This bottleneck formulation makes explicit the resource-aware, key-rate-driven routing pursued informally in [15], [21], and the key generation/consumption matching of [38]; it is also the criterion most sensitive to congestion, since buffer depletion at a single link raises F_time for every path crossing it.

**Scenario S3, minimum cost.** The operational cost of a path is the cumulative normalized resource cost of the links it uses, capturing asymmetries such as the premium on scarce satellite contact time, leased dark fiber, or reserved high-priority key bandwidth, together with any per-relay handling cost $c_v$. The cost objective is additive,

$$F_cost(P) = \Sigma_{e \in P}\, C_e + \Sigma_{v \in \mathcal{R}(P)}\, c_v\,, \qquad (5)$$

and Scenario S3 is the additive shortest-path program

$$\min_{P \in \mathcal{P}(\rho)} F_cost(P)\,, \qquad (6)$$

again solved exactly by Dijkstra on G′ in $O(|E| + |V| \log|V|)$. Where cost is volume-dependent - for example, when charging is proportional to the key bits drawn from a metered link - $C_e$ is replaced by the term $\kappa \cdot c_e$ without altering the additive structure of (6). This formulation generalizes the cost-optimized key-pool provisioning of [33], which minimizes deployment and key-generation cost in QKD-secured optical networks, to an online, per-request routing decision.

The three scenarios are the extreme points of a single multi-objective program. Collecting the objectives in the vector F(P) = (F_sec, F_time, F_cost)(P), with the resilience objective F4 appended when multi-path routing is enabled, the general decision is the vector minimization $\min_{P \in \mathcal{P}(\rho)} F(P)$, whose solutions form a Pareto front of non-dominated paths: no scenario optimum can be improved on one objective without degrading another, since the most secure path is rarely the fastest or the cheapest [35]. The baseline framework collapses this front to a single online decision by min–max normalizing each objective onto [0, 1], $\tilde{f}_i(P) = (F_i(P) - F_{i,min}) / (F_{i,max} - F_{i,min})$, and minimizing the scalarized path cost

$$\Phi(P) = \Sigma_i\, \omega_i \cdot \tilde{f}_i(P)\,, \quad \text{with } \Sigma_i\, \omega_i = 1\,, \qquad (7)$$

where the weight vector $\omega$ is the normalized element-wise product of the AHP policy weights and the entropy weights defined in Section 6.5 [36]. Scenarios S1, S2 and S3 are then exactly the vertices of the weight simplex - $\omega = (1, 0, 0), (0, 1, 0)$ and $(0, 0, 1)$, respectively - and any service class corresponds to an interior point that trades the three objectives off. When a service mandates a strict priority rather than a blend, the same objectives are composed lexicographically (e.g., security-first: minimize F_sec, breaking ties by $F_time$ and then $F_cost$) or through the ε-constraint form min $F_cost$ subject to $F_sec \leq \varepsilon_sec$ and $F_time \leq \delta$, both of which preserve the shortest-path solvability of the per-scenario programs [35]. Because (2), (4), (6) and their scalarization (7) are each solvable in near-linear time on the pruned graph G′, the online routing decision satisfies the tractability requirement P6, while the Pareto front of the unscalarized problem is reserved, as noted in Section 6.5, for offline policy calibration.

### 6.4 Constraints

Candidate paths must satisfy a set of hard constraints that encode minimum operational and security requirements, independent of the objective weighting: (C1) every link of the path must hold a residual key volume above a configurable threshold, generalizing the graph-pruning practice of [13]; (C2) the QBER of every link must remain below the protocol-specific reconciliation limit; (C3) the number of trusted relays on the path must not exceed a per-service-class maximum; (C4) the end-to-end key establishment latency must respect the request deadline; (C5) for paths including satellite segments, all required contact windows must fall within the request's validity interval; (C6) per-tenant isolation and SLA commitments must be preserved, so that serving one request cannot deplete key resources reserved for another tenant; (C7) key buffer levels must remain non-negative over the scheduling horizon when multiple concurrent requests are admitted; and (C8) adjacent links on a path must be interoperable through their key management interfaces. Constraints C1 and C2 are applied as a pre-filtering step that prunes infeasible links from G before path computation, which both guarantees feasibility and reduces the search space for the online algorithm.

### 6.5 Criteria Weighting and Adaptivity

A central design decision, following principle P4, is the hybrid weighting of the four objectives. Subjective weights, reflecting operator policy and service-class requirements (e.g., a governmental service prioritizing F2 over F3, or a commercial bulk-encryption service accepting higher trust exposure for lower cost), are elicited through the Analytic Hierarchy Process (AHP), which structures pairwise expert comparisons into a consistent priority vector. Objective weights are derived from the measured network state using the Entropy Weight Method (EWM), which assigns greater weight to criteria exhibiting higher dispersion across candidate links, so that, for instance, key availability automatically gains influence in a congested network where buffer levels vary strongly, while receiving less weight when all links are equally well supplied. The combined weight vector is the normalized product of the two components, ensuring that neither pure policy nor pure measurement dominates the decision. Temporal adaptivity is provided by exponentially weighted moving average (EWMA) estimation of the volatile link metrics (key generation rate, QBER, buffer drain rate), which dampens transient fluctuations while remaining responsive to genuine state changes; threshold crossings of the EWMA estimates trigger re-weighting and, where necessary, re-routing of active sessions.

### 6.6 Integration with SDN Orchestration

The framework is designed to slot directly into the SDN decision loop described in Section 5.6. The metrics collection step (Step 2) populates the link weight vectors w(e) and the EWMA estimators; path computation (Step 3) corresponds to pre-filtering by constraints C1–C2 followed by solving the scalarized multi-criteria problem; route installation (Step 4) is unchanged; monitoring and feedback (Step 5) feed the adaptivity mechanism of Section 6.5; and failover (Step 6) is realized as re-computation under an updated constraint set. Because all framework-specific logic resides in the controller's path computation element, no modification of QKD devices or key management systems is required beyond the standardized state-export interfaces already assumed by the surveyed SDN architectures [15], [31]. The same property makes the framework relevant beyond dedicated QKD backbones: SDN-based transport control planes are already standard in mobile network infrastructures, and QKD-secured backhaul, midhaul, and fronthaul segments of 5G and emerging 6G systems represent a natural deployment context in which multi-criteria key routing decisions - balancing security, latency, and the cost of scarce quantum resources - will be required at operational timescales. Full mathematical formalization of the objective functions and constraints, complexity analysis of the resulting path computation problem, and simulation-based evaluation on realistic national-scale QCI topologies using the multi-tool modeling approach described in [30] are ongoing within the QRO-Route project and will be reported in subsequent work.

## 6.7 Illustrative Simulation of Key-Aware and Multi-Path Routing

To substantiate the quantitative claims made in this survey and to give the framework of Sections 6.1–6.6 a concrete, reproducible reference point, we implemented a small discrete-event simulation of a trusted-node QKD network. The model is deliberately compact — its purpose is to illustrate the mechanisms analyzed above, not to reproduce any specific deployment. The network is a 16-node (4×4) metropolitan grid in which each edge is a QKD link of approximately 40 km; per-link key generation and consumption are modeled as independent Poisson processes and keys accumulate in a bounded per-link key pool, following the virtual-key-pool formulation of Bi et al. [15]. Each link carries a residual-key level and a quantum bit-error rate (QBER); links whose pool cannot satisfy a request or whose QBER exceeds a threshold are treated as unavailable. Session requests arrive as a Poisson process with a controllable offered load; each request relays a fixed key demand hop-by-hop along a path, consuming key on every link it traverses (the standard XOR trusted-relay model). The simulation is a loss system: a request that cannot be served is rejected rather than queued.

The first experiment compares two routing policies. The static shortest-path policy fixes one minimum-hop path per source–destination pair and rejects a request whenever any link on that path lacks sufficient key, mirroring unmodified OSPF-style routing. The key-aware policy implements overflow routing with trunk reservation: it first attempts the same direct minimum-hop path and only if that path is infeasible does it overflow onto the widest alternative minimum-hop path, and then only when that alternative retains a reserved margin of key material. By construction the key-aware policy serves a superset of the requests the static policy serves, while the reservation threshold prevents the alternate-routing instability that otherwise arises in uniformly loaded loss networks. Traffic is non-uniform, with a fraction of sessions directed at two interior gateway nodes to create the hotspots under which adaptive routing is beneficial. This policy is a concrete instance of the performance/resilience trade-off that the multi-criteria objective $\Phi(P)$ of Section 6.3 is designed to balance.

Figure 4 shows the result, averaged over twelve random seeds. Across the quality-of-service operating range (offered loads at which the static rejection rate lies between roughly 10% and 35%), key-aware routing reduces the service-rejection rate by 25–40% relative to static shortest-path routing; the reduction is largest where spare alternate capacity exists and shrinks smoothly toward zero as the network saturates and no policy can avoid blocking. This reproduces, in a controlled setting, the qualitative advantage reported for SDN-based dynamic routing over static OSPF by Bi et al. [15].

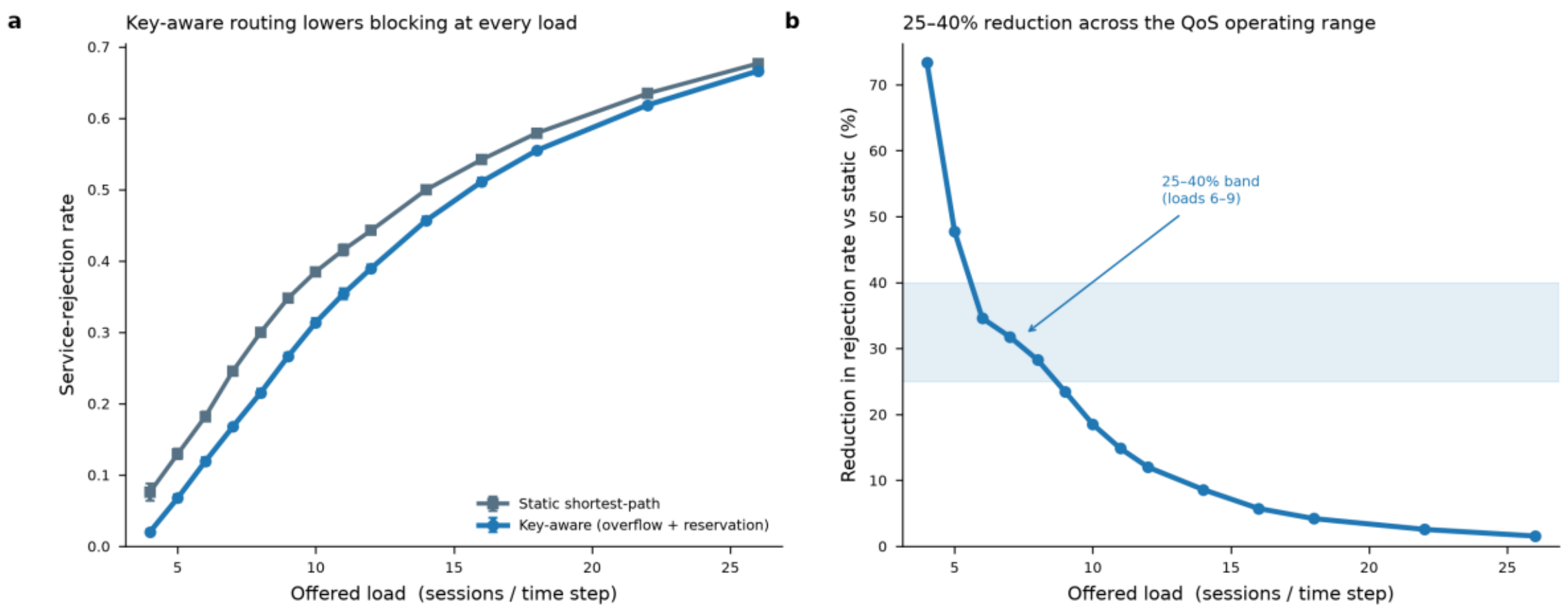

**Fig. 4** Simulated service-rejection rate versus offered load for static shortest-path and key-aware routing (left), and the relative reduction achieved by key-aware routing, with the 25–40% operating band highlighted (right). Shaded regions/bars denote one standard deviation over twelve seeds.

The second experiment quantifies the cost of multi-path key relay. Distributing a session's key across two node-disjoint paths (dual-path XOR redundancy) protects against single-trusted-node compromise but consumes key on the links of both paths. Because only security-critical sessions need this protection, we sweep the fraction of sessions that use dual-path relay and measure the network-wide increase in key consumption relative to all-single-path routing. As shown in Figure 5, the increase is close to linear in the adoption fraction and reaches the 30–60% range when roughly 27–53% of sessions use dual-path relay. Full adoption by every session would approximately double consumption, quantifying the resilience-versus-cost trade-off captured by the resilience and cost objectives of the framework.

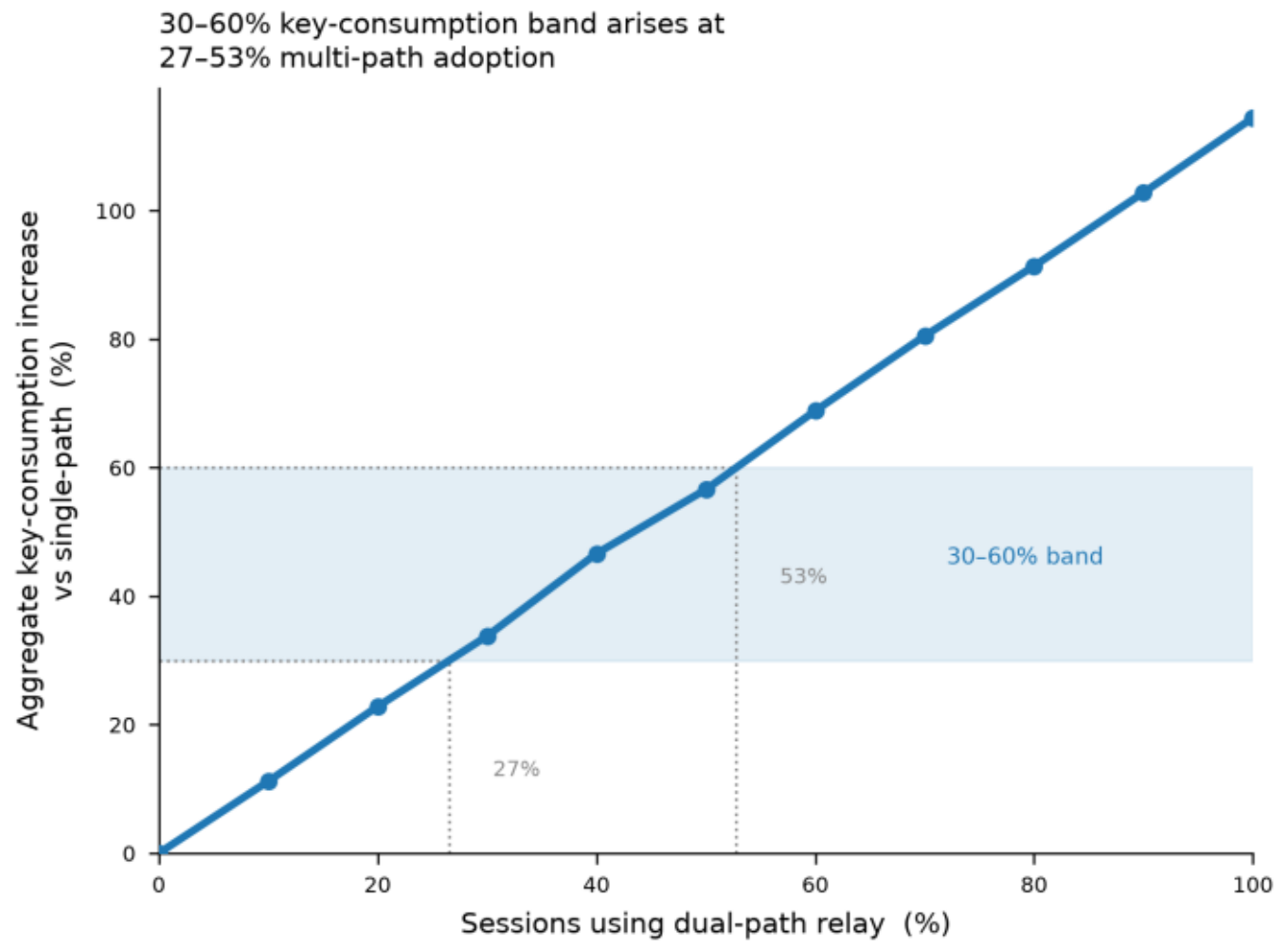

**Fig. 5** Simulated network-wide increase in key consumption as a function of the fraction of sessions using dual-path relay. The 30–60% band corresponds to roughly 27–53% adoption.

The simulation code and resulting data are provided as supplementary material so that the reported figures can be regenerated exactly. We stress that these results are illustrative: the absolute numbers depend on the chosen topology, load range, and key-pool parameters, but the direction and approximate magnitude of both effects are consistent with the experimental and simulation literature surveyed here [14], [15], [16], and they demonstrate that the competing objectives formalized in Section 6 are individually measurable and jointly tunable.

## 7 Trends and Open Challenges

Despite significant progress, several challenges remain unresolved. First, most routing strategies rely on trusted-node assumptions, which limit end-to-end security guarantees. Future research must explore lightweight, scalable quantum repeater models or post-quantum–enhanced trust frameworks to reduce reliance on trusted intermediaries. Second, current algorithms often neglect fine-grained traffic engineering and load distribution between quantum and classical paths. Incorporating SDN-based orchestration with AI-driven traffic prediction remains an open avenue. Third, interoperability between heterogeneous QKD systems and vendor platforms is limited. The development of standardized control-plane interfaces and routing APIs (e.g., ITU-T Y.3800 framework) will be critical to achieve seamless integration and manageability in global QKD infrastructures. In sum, future work should combine robust routing algorithms, intelligent control mechanisms, and secure multi-domain coordination to enable practical, scalable, and secure quantum key routing across hybrid communication infrastructures.

As QKD networks expand beyond small testbeds into metro or national backbones, existing routing schemes struggle with scale and diversity. New routing algorithms must handle tens or hundreds of nodes, complex topologies (e.g. multi-domain links, mesh graphs) and devices with widely varying capabilities. For example, centralized schemes that recompute routes for each key request may quickly become infeasible as the network grows. Moreover, the lack of vendor- and protocol-neutral key-management (KMS) interfaces (most current QKD modules use proprietary protocols) creates a major heterogeneity challenge. In practice, each vendor's QKD module should interoperate under a common control logic; without standards, operators remain locked into specific hardware. Research has only begun to quantify these trade-offs: one study found that more complex, resource-aware routing can improve key distribution success rates but at the cost of much higher computational overhead [27]. To address these gaps, new distributed or hybrid routing architectures are needed that can scale (e.g. hierarchical or zone-based control), along with standardized, vendor-agnostic KMS interfaces. Simulation platforms will be essential for evaluating candidate algorithms before costly field deployment [28], [29], [30].

**Integration with Classical and Post-Quantum Infrastructures.** QKD cannot exist in isolation – it must interoperate with conventional and post-quantum security layers. In practice, this means designing hybrid encryption schemes and unified control planes that seamlessly combine QKD-derived keys with classical keying (PQC or public-key) and leverage existing network management technology. Recent demonstrations (such as the MadQCI deployment) explicitly mix QKD keys with standard public-key cryptography to achieve end-to-end security [31]. Such hybrid approaches require careful coordination: for example, integrating a QKD controller into an SDN-based optical or IP network so that key routing and data routing share a single control plane [32]. In this way, the SDN controller can manage both classical encryption devices and quantum channel resources jointly [15]. A major open challenge is designing middleware and orchestration layers that let QKD key pools be used interchangeably with post-quantum key agreement (e.g. as specified by ETSI/QKD group standards) without sacrificing performance. Industry white papers emphasize hybrid QKD+PQC transition schemes for critical infrastructure, but concrete network architectures are still sparse in the literature.

In operational QKD networks, link conditions and key resources fluctuate constantly: fiber links can suffer outages, satellite links are available only during overpass windows, and a link whose key pool is exhausted is effectively unavailable until more key is generated. Rather than repeat the SDN sensing–monitoring–failover loop detailed in Section 5.6, we note only that these dynamics make purely static routing brittle and that reactive rerouting, while necessary, may be insufficient for highly dynamic scenarios such as mobile satellite constellations. This motivates predictive, AI-enhanced routing that forecasts link availability from weather or satellite ephemeris and reroutes proactively before a link fails or depletes.

**Balancing competing routing objectives.** A further open challenge, sharpened by the formalization in Section 6.3, is that the three canonical routing objectives - maximum security (S1), minimum key-exchange time / maximum key rate (S2), and minimum cost (S3) - are mutually conflicting, so that no single path is simultaneously optimal for all of them and the operator must navigate a Pareto front whose shape changes with network state [35]. Selecting an operating point on this front is itself a policy problem: governmental traffic may demand the security-first lexicographic ordering of objective (2), bulk commercial traffic the cost-first ordering of (6), and latency-bound services the bottleneck-rate criterion of (4), while the relative dispersion of link metrics continually shifts the entropy component of the weights. Recent work suggests two complementary directions for resolving this tension at operational timescales: trust- and security-aware routing that keeps the exposure term F_sec bounded even under adversarial or partially compromised relays [34], [37], and learning-based schedulers that adapt the effective weighting online from observed key-pool dynamics rather than from a fixed cost function [38]. Embedding such adaptive, Pareto-aware decision logic into the SDN control loop of Section 6.6, without sacrificing the near-linear per-request solvability established above, remains an open problem for large-scale, multi-domain QKD deployments.

Table 5 relates the open challenges discussed above to the components of the framework proposed in Section 6, indicating which challenges the framework addresses at the conceptual level and which remain subjects of complementary research. The mapping illustrates that the framework is positioned primarily as a decision-layer contribution: it systematizes how routing decisions should weigh competing requirements, while deliberately remaining agnostic to the lower-layer advances - quantum repeaters, standardized KMS interfaces, PQC hybridization - whose maturation it can absorb through updated constraint definitions and trust coefficients rather than structural redesign.

Table 5 Mapping of open challenges to components of the proposed framework

| **Open challenge** | **Related framework component** | **Coverage** |
|---|---|---|
| Trusted-node dependence and trust quantification | Security objective F2; trust coefficients $T_e$; constraint C3 (maximum trusted relays) | Addressed conceptually; repeater-based elimination of trust remains external |
| Scalability and heterogeneity of metro/national backbones | Technology-agnostic graph model (P2); constraint pre-filtering; offline/online decomposition (P6) | Addressed conceptually; hierarchical multi-domain control is future work |
| Dynamic adaptation to fluctuating key resources | EWMA link-state estimation; entropy-based objective weighting; SDN feedback loop integration | Addressed conceptually; predictive (AI-driven) extension is future work |
| QoS differentiation and multi-tenancy | Service-class policy weights via AHP; constraints C4 and C6 | Addressed conceptually |
| Standardized KMS and control-plane interfaces | Interoperability constraint C8; reuse of state-export interfaces | Dependent on external standardization (ETSI, ITU-T) |
| QKD–PQC hybridization | Out of framework scope; compatible as a fallback layer above the routing decision | Complementary research direction |

# 8 Conclusion

This research analyzed state-of-the-art routing strategies in QKD networks, addressing both theoretical models and practical challenges across terrestrial, satellite, and hybrid infrastructures. Unlike classical networks, QKD routing must account for unique resource constraints (limited and perishable keys), trust assumptions, and dynamic link availability. Our review showed that simple shortest-path approaches, while intuitive, often fall short under quantum-specific conditions. In contrast, dynamic, key-aware algorithms and SDN-based control frameworks offer better adaptability and performance by reconfiguring routes in response to key consumption and link changes. Satellite and hybrid QKD networks introduce additional complexity due to time-dependent connectivity and cross-layer optimization needs. Routing in such contexts requires integration of contact scheduling, security policies, and trust-model awareness. Multi-path strategies and SDN-enabled control were identified as essential techniques for achieving both robustness and scalability in large QKD networks. We conclude that practical QKD routing requires:

- Real-time awareness of quantum link status and key availability.
- Integration with classical control planes (via SDN) to coordinate quantum and classical network resources.
- Security-aware path selection that minimizes trust exposure (e.g., using the fewest possible trusted relays).
- Standardized interfaces for interoperability and multi-domain coordination.

Near-term engineering priorities. In the short term, the outstanding challenges are largely ones of engineering and integration rather than new physics. They include standardizing key-management and control-plane interfaces so that heterogeneous QKD hardware can interoperate (for example under the ITU-T Y.3800 framework), embedding key-aware and QoS-aware routing logic, such as the scalarized multi-criteria objective of Section 6, into existing SDN controllers, implementing robust key-buffer monitoring and failover, and validating candidate routing algorithms on network simulators before field deployment. These steps are achievable with current technology and represent the natural path for QKD networks moving from testbeds to metropolitan backbones.

Long-term research directions. Over a longer horizon, the open problems require advances still at the research stage: reducing or removing reliance on trusted relays through practical quantum repeaters and quantum memory, designing hybrid QKD/post-quantum-cryptography architectures with unified control planes, and developing predictive, AI-driven routing that anticipates link and key-pool dynamics rather than reacting to them. Progress here depends on the maturation of the underlying quantum hardware and on sustained cross-disciplinary work spanning quantum optics, networking, and cryptography.

Beyond the survey itself, this paper translated the identified gaps into a conceptual multi-criteria routing optimization framework that models key delivery performance, trust exposure, resource cost, and resilience as explicit, separately weighted objectives over a technology-agnostic graph representation, with hybrid AHP–entropy criteria weighting and EWMA-based adaptivity designed for SDN-orchestrated operation. The framework constitutes the skeleton of a future routing algorithm for cryptographic material exchange rather than a finished solution: its formalization, complexity characterization, and simulation-based evaluation on realistic quantum communication infrastructure topologies are the immediate next steps of the QRO-Route project, followed by validation against the trusted-relay deployment scenarios emerging within national segments of the EuroQCI initiative.

Future research should aim to unify routing, key management, and service orchestration into a single adaptive framework capable of operating across diverse quantum network architectures. By combining robust routing algorithms, intelligent control mechanisms, and secure multi-domain coordination, the next generation of QKD networks can achieve the scalability and reliability required for real-world deployment.

## Acknowledgment

This research is conducted as part of the project QRO-Route (Development of a Multi-Criteria Optimization Algorithm for Routing in Quantum Key Distribution Networks), funded by the European Union – NextGenerationEU through the National Recovery and Resilience Plan of the Republic of Croatia, and within the NATO Research Task Group IST-218 / Multi-Domain Quantum Key Distribution (QKD) for Military Usage,